\documentclass{emulateapj}
\usepackage{epsfig,natbib}
\usepackage{graphicx}
\newcommand{\ms}{\mbox{m\,s$^{-1}~$}}

\newcommand{\kse}{\mbox{km\,s$^{-1}$}}
\newcommand{\mse}{\mbox{m\,s$^{-1}$}}

\newcommand{\msye}{\mbox{m\,s$^{-1}$\,yr$^{-1}$}}

\newcommand{\mearth}{$M_\earth$~}
\newcommand{\mearthe}{$M_\earth$}
\newcommand{\rearth}{$R_\earth$~}

\newcommand{\msinie}{$M \sin i$}

\newcommand{\chinu}{$\chi_{\nu}$}

\newcommand{\feh}{\ensuremath{[\mbox{Fe}/\mbox{H}]}}
\newcommand{\rphk}{\ensuremath{R'_{\mbox{\scriptsize HK}}}}
\newcommand{\lrphk}{\ensuremath{\log{\rphk}}}
\newcommand{\caii}{\ion{Ca}{2} H \& K}
\newcommand{\caiih}{\ion{Ca}{2} H}

\newcommand{\etaearth}{$\mathbf \eta_{\oplus} ~$}
\newcommand{\etaearthe}{$\mathbf \eta_{\oplus}$}

\slugcomment{Received by ApJ 2008 Oct 31; accepted 2009 Jan 26}

\shortauthors{Howard {et~al.}}
\shorttitle{A Super-Earth Orbiting HD\,7924}
\begin{document}
\pagenumbering{arabic}


\title{The NASA-UC Eta-Earth Program: I. A Super-Earth Orbiting HD\,7924\altaffilmark{1}}
\author{
Andrew W.\ Howard\altaffilmark{2,3}, 
John Asher Johnson\altaffilmark{4}, 
Geoffrey W.\ Marcy\altaffilmark{2}, 
Debra A.\ Fischer\altaffilmark{5}, \\
Jason T.\ Wright\altaffilmark{6}, 
Gregory W.\ Henry\altaffilmark{7},
Matthew J.\ Giguere\altaffilmark{5}, 
Howard Isaacson\altaffilmark{5}, \\
Jeff A.\ Valenti\altaffilmark{8},
Jay Anderson\altaffilmark{8}, and
Nikolai E.\ Piskunov\altaffilmark{9} 
}
\altaffiltext{1}{Based on observations obtained at the W.\,M.\,Keck Observatory, 
                      which is operated jointly by the University of California and the 
                      California Institute of Technology.  Keck time has been granted by both 
                      NASA and the University of California.} 
\altaffiltext{2}{Department of Astronomy, University of California, Berkeley, CA 94720-3411 USA} 
\altaffiltext{3}{Townes Fellow, Space Sciences Laboratory, University of California, 
                        Berkeley, CA 94720-7450 USA; howard@astro.berkeley.edu}
\altaffiltext{4}{Institute for Astronomy, University of Hawaii, 
                        Honolulu, HI 96822 USA; NSF Postdoctoral Fellow}
\altaffiltext{5}{Department of Physics and Astronomy, San Francisco State University, 
                        San Francisco, CA 94132 USA}
\altaffiltext{6}{Department of Astronomy, Cornell University, Ithaca, NY 14853 USA}
\altaffiltext{7}{Center of Excellence in Information Systems, Tennessee State University, 
                        3500 John A.\ Merritt Blvd., Box 9501, Nashville, TN 37209 USA}
\altaffiltext{8}{Space Telescope Science Institute, 3700 San Martin Dr., Baltimore, MD 21218, USA}
\altaffiltext{9}{Department of Astronomy and Space Physics, Uppsala University, 
                        Box 515, 751 20 Uppsala, Sweden}

\begin{abstract}
We report the discovery of the first low-mass planet to emerge from the NASA-UC Eta-Earth Program, 
a super-Earth orbiting the K0 dwarf HD\,7924.  
Keplerian modeling of precise Doppler radial velocities reveals a planet with 
minimum mass $M_P\sin i$\,=\,9.26\,\mearth in a $P$\,=\,5.398\,d orbit.
Based on Keck-HIRES measurements from 2001 to 2008, 
the planet is robustly detected with an estimated false alarm probability of less than 0.001.   
Photometric observations using the Automated Photometric Telescopes at Fairborn Observatory show that 
HD\,7924 is photometrically constant over the radial velocity period to 
0.19\,mmag, supporting the existence of the planetary companion.
No transits were detected down to a photometric limit of $\sim$0.5\,mmag, 
eliminating transiting planets with a variety of compositions.
HD\,7924b is one of only eight planets known with $M_P\sin i$\,$<$\,10\,\mearth 
and as such is a member of an 
emerging family of low-mass planets that together constrain theories of planet formation.
\end{abstract}



\section{Introduction}
\label{sec:intro}

The NASA-UC Eta-Earth Survey by the California Planet Search (CPS) group 
is a systematic search for low-mass planets 
($\sim$3--30\,\mearthe) orbiting the nearest 230 GKM stars 
suitable for high-precision Doppler observations at the Keck Observatory.  
In order to place statistically significant constraints on the fraction of stars with  
Earth-mass planets in the habitable zone, 
\etaearthe, and theories of planet formation, 
each star is observed a minimum of 20 times at $\sim$1\,\ms precision, 
including at least one set of high-cadence observations on consecutive nights.  
Planet candidates are followed up with additional high-cadence observations 
to confirm the candidate signal 
and to accurately measure the planet's orbital parameters. 
While Doppler surveys cannot detect 1\,\mearth planets in the habitable zone, 
we will estimate \etaearth by modest extrapolation from the distribution of detected 
super-Earths and Neptune-mass planets in shorter-period orbits.  
This paper is the first in a series to describe the planets and planetary systems 
emerging from the Eta-Earth Program.  

The statistics of planet occurance from the Eta-Earth Survey will offer 
important constraints on competing theories of planet formation 
\citep{Ida04a,Kenyon06,Alibert05,Mordasini07,Ida_Lin08_iv}.
These models differ in assumptions about the growth rate of dust into planetesimals, 
the viscosity of the disk, the location and effects of the snow line in the disk, 
the efficacy of inward migration, the accretion of gas and water, 
and the relevance of planet-planet interactions.
They are consistent with detections and measurements of jovian gas giants 
\citep[e.g.][]{Marcy_Japan_05,Udry2003}.
However, these theories predict that planets of mass 1--30\,\mearth are rare within 1\,AU, 
forming a ``planet desert'' of super-Earths 
($M_{\mathrm{pl}}$\,sin\,$i$ $\le$ 10\,\mearthe) and Neptune-mass planets.  
All models predict that Type I migration quickly ($\tau$\,$\sim$\,10$^5$\,yr) 
causes the rocky planets to spiral inward, destined to be lost in the 
star\footnote{While some of these models (e.g.\ \citealt{Ida_Lin08_iv}) 
show a pile-up of low-mass planets in short-period orbits
(analogous to hot Jupiters), this concentration may be artificial since the physical mechanism 
for stopping the migration remains uncertain.}.
Meanwhile, the more massive rocky cores ($M$\,$>$\,1\,\mearthe) accrete gas quickly, 
becoming ice- or gas-giants.
The resulting distribution of planetary systems is depleted 
of planets in the mass range 1--30\,\mearth within 1\,AU.

If the prediction of a low-mass desert is contradicted by 
a statistically well-defined sample of stars (such as the Eta-Earth Survey),  
then planet formation theory must be significantly modified with new physics.   
The planet desert has indeed been challenged by 
\citet{Mayor09} who estimate that 30\%\,$\pm$\,10\% of GK dwarfs 
have rocky or Neptune-mass planets inward of 50\,d orbits.
This claim is based in part on detections by the Swiss group of 
three Neptune-size planets orbiting HD\,69830 \citep{Lovis2006}, 
two of three planets orbiting GJ\,581 in or near the Habitable Zone \citep{Bonfils2005, Udry2007}, 
and a triple super-Earth system around HD\,40307 \citep{Mayor09}.

It is against this backdrop of recent discoveries and competing claims of the abundance 
of rocky planets that we announce the detection of a super-Earth 
orbiting the K dwarf HD\,7924.  
\S\,\ref{sec:props} of this paper describes the host star, HD\,7924.  
The spectroscopic observations and their Doppler reduction are described in \S\,\ref{sec:obs}.
Our detection of HD\,7924b, 
including Keplerian fitting and false alarm probability estimation, 
is described in \S\,\ref{sec:orbital}.  
In \S\,\ref{sec:second_planet},
we consider and ultimately reject a second planetary companion based on the current data.
Our photometric observations and their constraints on planetary transits 
are described in  \S\,\ref{sec:photometry}.
We conclude in \S\,\ref{sec:discussion} with a discussion of HD\,7924b and the emerging 
family of super-Earth planets.

\begin{deluxetable}{lc}
\tablecaption{Stellar Properties of HD\,7924
\label{tab:stellar_params}}
\tablewidth{0pt}
\tablehead{
  \colhead{Parameter}   & 
  \colhead{Value} 
}
\startdata
Spectral type ~~~~~~~~~~~~~~~~& K0\,V\\
$M_V$ & 6.056\\
$B-V$ & 0.826\\
$V$   & 7.185\\
Distance (pc) & 16.8\,$\pm$\,0.1\\
\feh& $-0.15$\,$\pm$\,0.03\\
$T_\mathrm{eff}$ (K) &  5177\,$\pm$\,44\\
$v$\,sin\,$i$ (km\,s$^{-1}$) & 1.35\,$\pm$\,0.5 \\
log\,$g$ & 4.59$^{+0.02}_{-0.03}$ \\
$L_{\star}$ ($L_{\sun}$) & 0.368$^{+0.019}_{-0.018}$\\
$M_{\star}$ ($M_{\sun}$) & 0.832$^{+0.022}_{-0.036}$\\
$R_{\star}$ ($R_{\sun}$) & 0.78\,$\pm$\,0.02\\
$S_\mathrm{HK}$ & 0.20\\
\lrphk & $-4.89$\\
$P_\mathrm{rot}$ (days) & $\sim$38\\
$\sigma_\mathrm{phot}$ (mag) & $<$\,0.0017\\
\enddata
\end{deluxetable}

\section{Properties of HD\,7924}
\label{sec:props}

HD\,7924 (HIP\,6379, GJ\,56.5) is a K0 dwarf whose properties 
are summarized in Table \ref{tab:stellar_params}.
It is nearby, $d$\,=\,16.8\,pc, and relatively bright, $V$\,=\,7.185 \citep{Perryman97,vanLeeuwen2007}.
With $M_V$\,=\,6.056 and $B-V$\,=\,0.826, HD\,7924 star lies 0.27\,mag  
below the $Hipparcos$ average main sequence as defined by \citet{Wright05}.

Using the SME (Spectroscopy Made Easy) LTE spectral synthesis code, 
\citet{Valenti05} analyzed a high-resolution, iodine-free 
Keck-HIRES spectrum of HD\,7924 and found 
the effective temperature $T_{\mathrm{eff}}$\,=\,5177\,$\pm$\,30\,K, 
surface gravity log\,$g$\,=\,4.58\,$\pm$\,0.08, 
projected rotational velocity $v\sin i$\,=\,1.35\,$\pm$\,0.5\,\kse, 
and iron abundance ratio [Fe/H]\,=\,$-0.15$\,$\pm$\,0.03 
(i.e.\ slightly metal poor, consistent with its location below the average main sequence).
Using these quantities and catalog values for $V$-band photometry, parallax, etc., 
they interpolated the Yonsei-Yale isochrones to obtain the 
luminosity $L_{\star}$\,=\,0.368$^{+0.019}_{-0.018}$\,$L_{\sun}$, 
radius $R_{\star}$\,=\,0.754\,$\pm$\,0.013\,$R_{\sun}$, and 
mass $M_{\star}$\,=\,0.80$^{+0.04}_{-0.03}$\,$M_{\sun}$.  
Isochrones provide no useful constraint on the age because the star 
has not begun evolving off the main sequence. 

\citet{Takeda07} used the spectroscopically determined stellar parameters from 
\citet{Valenti05} and derived various stellar parameters 
by matching those parameters to theoretical stellar evolution models.
Their values are largely consistent with \citet{Valenti05}: 
$M_{\star}$\,=\,0.832\,$^{+0.022}_{-0.036}$\,$M_{\sun}$, 
$R_{\star}$\,=\,0.78\,$\pm$\,0.02\,$R_{\sun}$, and 
log\,$g$\,=\,4.59\,$^{+0.02}_{-0.03}$.
We adopt the Takeda et al.\ values for the calculations below.

\begin{figure}[t]
\vspace*{0.45in}
\hspace*{0.25in}
\includegraphics[width=0.42\textwidth]{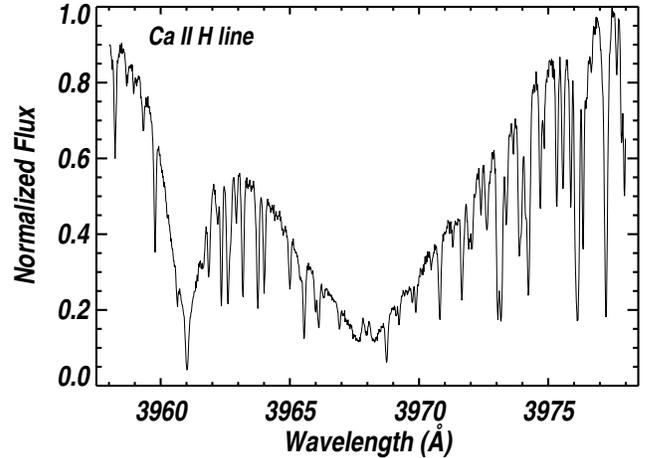}
\caption{\caiih\ line for HD\,7924.  
Slight line core emission is seen indicating modest chromospheric activity.}
\label{fig:caii}
\end{figure}

Measurements of the cores of the \caii\ lines show that 
HD\,7924 has modest chromospheric activity (Fig.\ \ref{fig:caii}).
We measured the chromospheric activity indices 
$S_{HK}$\,=\,0.20 and \lrphk\,=\,$-$4.89  \citep{Wright04}.
The full set of $S_{HK}$ measurements for all observations of HD\,7924 is nearly constant in time 
(fractional rms of 1.2\%) 
and does not show a periodicity at or near 5.4\,d (the planet's orbital period; see \S\,\ref{sec:orbital}).
We estimate $P_{\mathrm{rot}}$\,=\,38\,d and an age of 3.8\,Gyr
using \rphk\ and $B-V$ calibration \citep{Wright04}.
Following \citet{Wright05}, and based on the values of $S_{HK}$, $M_V$, and $B-V$, 
we estimate a radial velocity (RV) jitter of 2.13\,\mse.
This empirical estimate for jitter accounts for RV variability due to 
rotational modulation of stellar surface features, stellar pulsation, undetected planets, 
and uncorrected systematic errors in the velocity reduction \citep{Saar98,Wright05}. 
As explained in \S\,\ref{sec:orbital}, jitter is added in quadrature to the 
RV measurement uncertainties for Keplerian fitting.

\section{Observations and Doppler Reduction}
\label{sec:obs}

We observed HD\,7924 using the HIRES echelle spectrometer \citep{Vogt94} 
on the 10-m Keck I telescope.  The 198 observations span eight years (2001--2008) 
with high-cadence observations---clusters of observations on 6--12 
consecutive nights---beginning in late 2005.  
All observations were made with an iodine cell mounted directly in front of the 
spectrometer entrance slit.  The dense set of molecular absorption lines imprinted 
on the stellar spectra provide a robust wavelength fiducial 
against which Doppler shifts are measured, 
as well as strong constraints on the shape of the spectrometer instrumental profile at 
the time of each observation \citep{Marcy92,Valenti95}.

We measured the Doppler shift from each star-times-iodine spectrum using a 
modeling procedure modified from the method described by \citet{Butler96b}.  
The most significant modification is the way we model the
intrinsic stellar spectrum, which serves as a reference point for the relative Doppler
shift measurements for each observation.  Butler et al.\ use a version of the
\citet{Jansson95} deconvolution algorithm to remove the spectrometer's instrumental profile
from an iodine-free template spectrum.  We instead use a new deconvolution
algorithm developed by one of us (J.\,A.\,J.) that employs a more effective regularization
scheme, which results in significantly less noise amplification and improved
Doppler precision. We defer a more detailed description of this method to a
forthcoming publication (Johnson et al. 2009, in prep).

\begin{figure}
\vspace*{0.35in}
\hspace*{-0.5in}
\plotone{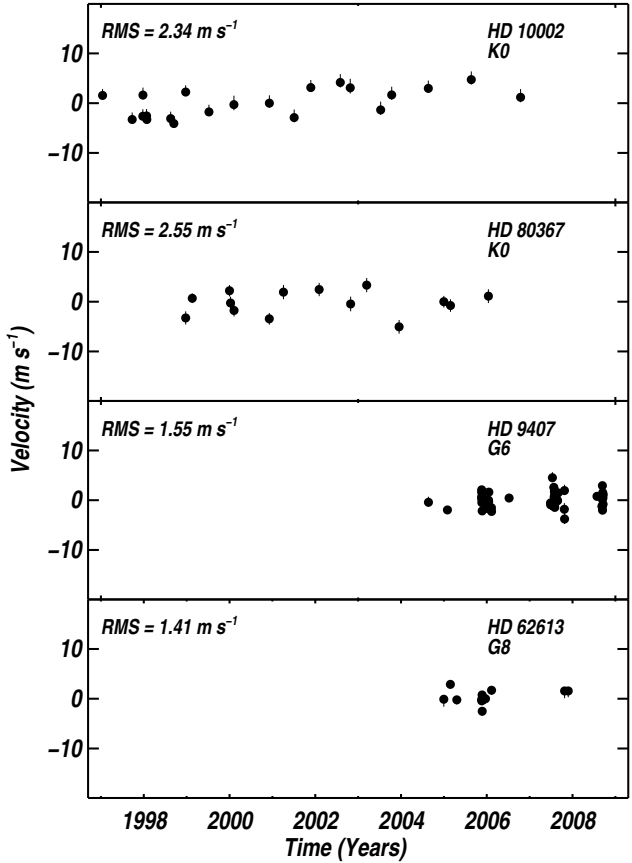}
\caption{Radial velocity time series for four stable stars  
in our Keck Doppler survey that are similar to HD\,7924.
These stars demonstrate long-term velocity stability and  
improved Doppler performance for observations after the HIRES upgrade in 2004.
The bottom two panels show only post-upgrade measurements.  
The binned velocities with measurement uncertainties (but not jitter) are plotted.  
Panels are labeled with star name, spectral type, and reduced rms to a linear fit.}
\label{fig:standard_stars}
\end{figure}

Figure \ref{fig:standard_stars} shows radial velocity time series for four stable stars with
characteristics similar to HD\,7924, demonstrating our measurement precision over 
the past 4--11 years. 
In 2004 August, the Keck HIRES spectrometer was upgraded with a new detector. 
The previous 2K\,$\times$\,2K pixel Tektronix CCD was replaced by an array of 
three 4K\,$\times$\,2K pixel MIT-LL CCDs. 
The new detector produces significantly higher velocity precision due to its improved 
charge transfer efficiency and charge diffusion characteristics, smaller pixels  
(15\,$\mu$m vs.\ 24\,$\mu$m), higher quantum efficiency, 
increased spectral coverage, and lower read noise.
Our post-upgrade measurements exhibit a typical long-term rms scatter of $\sim$1.5\,\mse,
compared to $\sim$2.5\,\ms for pre-upgrade measurements\footnote{
Measurements prior to JD 2,453,237 are pre-upgrade.}.
The pre- and post-upgrade measurements also lack a common velocity zero point, 
but we've fit for and corrected this offset to within $\sim$2\,\ms 
for every star observed at Keck using a large set of 
stable stars with many pre- and post-upgrade observations.
To further limit the impact of the velocity discontinuity for this star, 
we let the offset float in the Keplerian fits below, 
effectively treating pre-upgrade and post-upgrade observations as coming 
from two different telescopes.  

\begin{figure}
\vspace*{0.25in}
\hspace*{0.25in}
\plotone{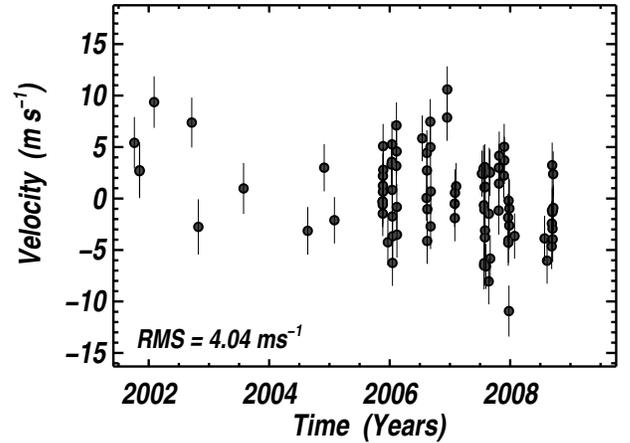}
\caption{Time series of binned radial velocities of HD\,7924 from
Keck-HIRES with error bars indicating the quadrature sum of 
measurement uncertainties and jitter.}
\label{fig:time_series}
\end{figure}

The velocities derived from the 198 observations have 
a median measurement uncertainty of 0.9\,\ms for single measurements.  
This uncertainty is the weighted standard deviation of the mean of the velocity measured 
from each of the $\sim$700 2\,$\mathrm{\AA}$ chunks in each echelle spectrum \citep{Butler96b}.
In many cases, we made consecutive observations of HD\,7924 to reduce the 
Poisson noise from photon statistics.  
For the Keplerian analysis below (\S\,\ref{sec:orbital}--\ref{sec:second_planet}), 
the velocities were binned in 2\,hr intervals 
yielding 93 measurements with an rms of 4.04\,\ms about the mean and 
a median measurement uncertainty of 0.7\,\mse.  
The binned velocities and associated measurement 
uncertainties are listed in Table \ref{tab:keck_vels} [at the end of this preprint, after References] 
and plotted as a time series in Fig.\ \ref{fig:time_series}.

\section{Orbital solution}
\label{sec:orbital}

We searched the radial velocities in Table \ref{tab:keck_vels} for the best-fit Keplerian 
orbital solution using the partially-linearized, least-squares fitting procedure described 
in \citet{Wright08}.  
Each velocity measurement was assigned a weight 
constructed from the quadrature sum of the measurement uncertainty 
(listed in Table \ref{tab:keck_vels}) and a stellar jitter term (2.13\,\mse).  

\begin{figure}
\vspace*{0.25in}
\hspace*{0.2in}
\plotone{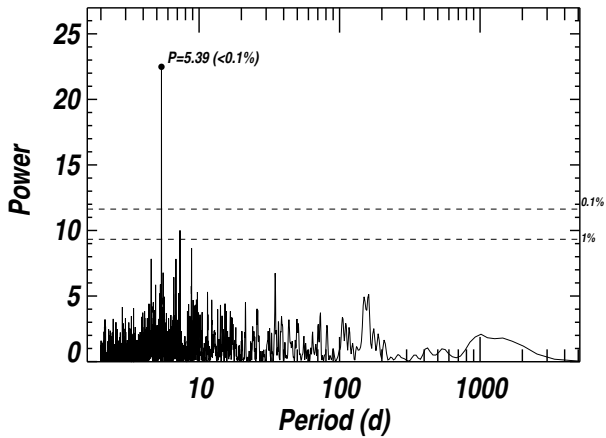}
\caption{Lomb-Scargle periodogram of binned radial velocity measurements of HD\,7924.
The tall peak near $P=5.39$\,d suggests a planet with that orbital period.}
\label{fig:pergram}
\end{figure}

\begin{figure}
\vspace*{0.2in}
\hspace*{0.1in}
\plotone{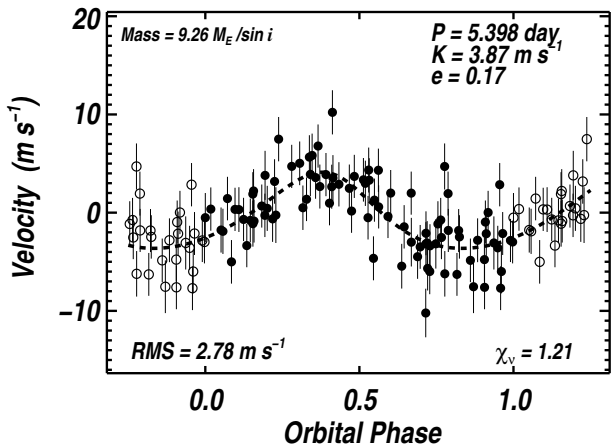}
\caption{Single-planet Keplerian model for the radial velocities of HD\,7924, 
as measured by Keck-HIRES.  
The dashed line shows the best-fit Keplerian orbital solution 
representing a 9.26\,M$_\earth$ (minimum mass) planet in a 5.398\,d orbit.  
Filled circles represent phased and binned radial velocities, 
while the open circles show the same velocities wrapped one orbital phase.
The error bars show the quadrature sum of measurement uncertainties 
and 2.13\,\ms jitter.}
\label{fig:phased_1p}
\end{figure}

The Lomb-Scargle periodogram of the velocities (Fig.\ \ref{fig:pergram}) shows
a strong periodic signal near 5.39\,d.  
We seeded the Keplerian search with this period, 
and a wide variety of other trial periods, and found the best-fit orbital solution shown 
in Fig.\ \ref{fig:phased_1p}.  This minimum in $\chi_{\nu} = \sqrt{\chi_{\nu}^2}$ 
implies a planet in a 5.3978\,$\pm$\,0.0015\,d orbit around HD\,7924 
with a Doppler semi-amplitude of 3.87\,$\pm$\,0.72\,\ms 
and an orbital eccentricity of 0.17\,$\pm$\,0.16.  
Note that the orbit is consistent with circular, consistent with 
the well-known bias against measuring $e$\,=\,0 in  
low amplitude systems (e.g., \citealt{OToole2009}).
The best-fit orbit includes a linear trend, $dv/dt$\,=\,$-1.07$\,$\pm$\,0.35\,\msye.
The velocity offset between pre- and post-upgrade HIRES observations was solved for 
and is incorporated in Table \ref{tab:keck_vels}.
Using our adopted stellar mass of 0.832\,$M_{\sun}$, we find a 
minimum planet mass of $M_P\sin i$\,=\,9.26\,$M_\earth$ 
and an orbital semi-major axis of $a$\,=\,0.057\,AU.  
The parameters of this orbital solution, and associated statistics, 
are listed in Table \ref{tab:orbital_params}.

\begin{deluxetable}{lc}
\tabletypesize{\footnotesize}
\tablecaption{Orbital Solution for HD\,7924b
\label{tab:orbital_params}}
\tablewidth{0pt}
\tablehead{
\colhead{Parameter}   & \colhead{Value} 
}
\startdata
$P$ (days)     & 5.3978 $\pm$ 0.0015 \\
$e$                     & 0.17 $\pm$ 0.16 \\
$K$ (m\,s$^{-1}$)       & 3.87 $\pm$ 0.72 \\
$T_p$ (JD -- 2,440,000) & 14,727.27 $\pm$ 0.87 \\
$\omega$ (deg)          & 25 $\pm$ 60 \\
$dv/dt$ (\msye) &$-1.07$\,$\pm$\,0.35 \\
$M$\,sin\,$i$ ($M_\earth$) & 9.26\\
$a$ (AU)                & 0.057\\
$N_\mathrm{obs}$ (binned) & 93 \\
Median binned uncertainty (\mse) & 0.7 \\
Assumed jitter (\mse) & 2.13 \\
Reduced rms to fit (\mse) & 2.78 \\
$\sqrt{\chi^2_\nu}$  & 1.21 \\
FAP (periodogram peak) &  $<$0.0001\\
FAP (\chinu\ for Keplerian fit) &  $<$0.001\\
\enddata
\end{deluxetable}

The Keplerian parameter uncertainties were derived using a Monte Carlo 
method \citep{Marcy05} and do not account for correlations between parameter errors.  
With seven years of observations of a 5\,d planet, the error on $P$ is quite small 
(one part in 3600) and largely uncorrelated with the errors on other Keplerian parameters.  
But because $K$ is relatively small, the data are nearly compatible with a family of 
Keplerians ranging from circular to slightly eccentric, 
leading to fractionally higher, correlated errors on the other parameters 
(especially $e$ and $\omega$).
Our estimate for a possible time of mid-transit (see \S\,\ref{sec:photometry}) 
accounts for these correlated parameters, 
thereby reclaiming some of the clock-like precision of the period estimate.

As we show in \S\,\ref{sec:photometry}, HD\,7924 is photometrically stable, particularly on 
the short time scales near the 5.4\,d Keplerian period.  
Thus, the radial velocity signal we detected is almost certainly not due to  
rotational modulation of spots or other stellar surface features.

We considered the possibility that the velocity periodicity represented by the 
Keplerian orbital fit in Fig.\ \ref{fig:phased_1p} arose from chance fluctuations in the velocities.
The amplitude of our orbital solution is less than 4\,\mse, 
which is within a factor of two of the quadrature sum of measurements uncertainties and jitter, 
so a careful treatment is deserved.  
We tested the null hypothesis in many ways, including several estimates of the 
false alarm probability (FAP).  

First, we note that the unreduced value of $\chi^2$ decreased from 255 for a linear 
fit to the data ($N_{\mathrm{obs}} - 2 = 91$ degrees of freedom) to 125 for 
the single-planet Keplerian fit (85 degrees of freedom).
This corresponds to a decrease in $\chi_{\nu}$ from 1.68 to 1.21, and 
a decrease in the reduced velocity rms from 3.81 to 2.78\,\mse.  

\begin{figure}
\vspace*{0.25in}
\hspace*{0.1in}
\plotone{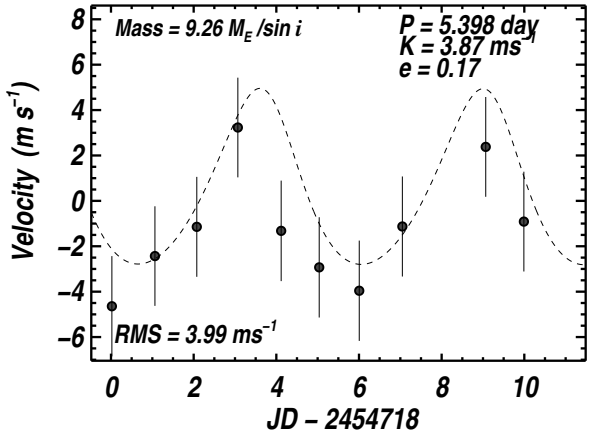}
\caption{High-cadence radial velocities of HD\,7924 during a long observing run.
Measurements were made on 10 of 11 consecutive nights.  
The velocities show temporal coherence over two orbital phases that matches  
the best-fit Keplerian orbital solution to all of the velocities for HD\,7924 
(the same orbital solution as in Fig.\ \ref{fig:phased_1p}).
The error bars show the quadrature sum of measurement uncertainties 
and 2.13\,\ms jitter.}
\label{fig:time_series_j59}
\end{figure}

Figure \ref{fig:time_series_j59} shows that all of the velocities measured during 
a recent 11-night observing run appear consistent in period, phase, and amplitude with  
the overplotted best-fit Keplerian solution that was derived from the complete set of 93 velocities.

\begin{figure}
\vspace*{0.2in}
\hspace*{0.15in}
\plotone{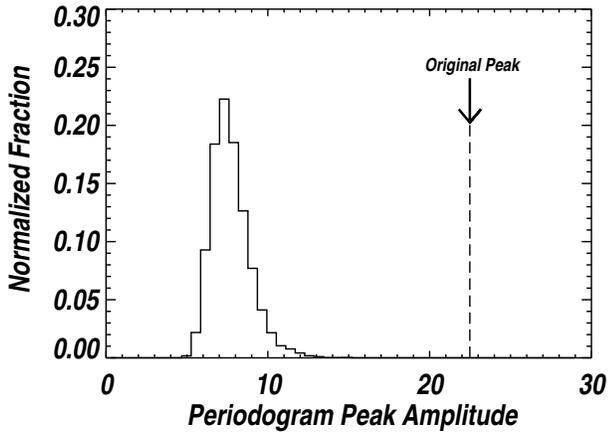}
\caption{FAP determination for HD\,7924 using periodogram peak height.  
The histogram shows the peak amplitudes from the periodograms of $10^4$ 
synthetic RV data sets, selected with replacement from the measured RVs.
None of the trials yielded a periodogram peak amplitude greater than the original fit, 
implying FAP\,$<$\,$10^{-4}$.}
\label{fig:fap_per}
\end{figure}

We also computed multiple false alarm probabilities associated with the 
chance arrangement of random velocities masquerading as a coherent 
signal \citep{Marcy05,Cumming04}.  
Figure \ref{fig:fap_per} shows the periodogram-based FAP test.  
In this test, we created $10^4$ synthetic data sets by drawing (with replacement) 
the velocities and associated errors from the set of measured velocities and using 
the actual observation times.  
A periodogram was computed for each synthetic data set and the peak amplitudes 
(typically in the range $\sim$5--10) are plotted as a histogram in Fig.\ \ref{fig:fap_per}.
This distribution is cleanly separated from the periodogram peak amplitude of the 
measured velocities, implying a strong coherence in the measured velocities 
well in excess of random fluctuations.  
We conclude that it is extremely unlikely (FAP $\ll$ $10^{-4}$) 
that a rearrangement of the measured velocities would result in a 
Fourier decomposition with as much power in a single mode as we observed.  
That is, a circular or nearly circular orbit of equal or greater amplitude
would almost never be falsely detected.

\begin{figure}
\vspace*{0.25in}
\hspace*{0.15in}
\plotone{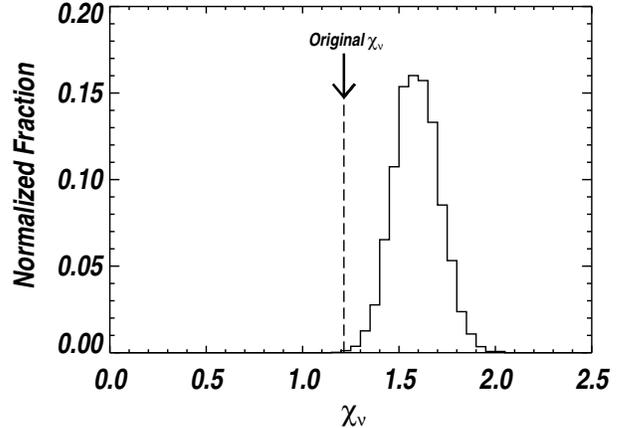}
\caption{FAP determination for HD\,7924 using \chinu.  
The histogram shows \chinu\ from the best-fit Keplerian solution to 1000 
synthetic RV data sets, selected with replacement from the measured RVs.
None of the trials yielded a \chinu\ lower than the original fit, implying FAP\,$<$\,0.001.}
\label{fig:fap_chi}
\end{figure}

While instructive, the periodogram-based test 
underestimates the true FAP associated with searching for a Keplerian signal because 
it only measures the strength of sinusoidal signals.  
To overcome this limitation 
we use a test based on \chinu\ for a full Keplerian fit, instead of periodogram peak height.
As with the previous test, the synthetic data sets were constructed  
by drawing with replacement from the measured velocities.  
Figure \ref{fig:fap_chi} shows the distribution of \chinu\ 
for the synthetic data sets as well as the value of \chinu\ for 
the Keplerian fit to the original, unscrambled velocities.  
None of the synthetic data sets had a best-fit Keplerian with 
\chinu\ lower than for the unscrambled velocities, 
implying an FAP for this scenario of $<$\,0.001.    

This FAP based on \chinu\ is also instructive, but is probably 
too conservative for at least two reasons.
(Note that these caveats are irrelevant for HD\,7924b since we have already shown that the FAP 
is very small, but are important for the possible second companion 
considered in \S\,\ref{sec:second_planet} and for future planetary systems.) 
First, it considers all Keplerian solutions equally believable, 
independent of orbital eccentricity.  
To be sure, nature occasionally produces planetary systems with eccentricities of $\sim$0.9 
(e.g.\ HD\,80606; \citealt{Naef2001}) that would have seemed unbelievable not long ago.
Yet, when these models arise as solutions to synthetic data sets, 
the high eccentricity and low \chinu\ are often driven by a 
few aberrant points that sometimes arise in a data set created from non-Gaussian noise
drawn with replacement.  
When such solutions are proposed for real data, they are often regarded with skepticism 
until the eccentric Keplerian hypothesis can be tested with 
additional measurements around the time of periastron passage. 
Thus, in accommodating all Keplerian signals, this test overestimates the FAP.

The second reason the \chinu-based FAP is sometimes too conservative 
is also a failure of over-accommodation.   
This test (as well as the periodogram-based FAP test) assumes  
that the observed velocity variation is \textit{completely} due to noise. 
However we have independent estimates of the expected velocity variability of a given star 
from the Doppler analysis pipeline, activity indices, and comparison stars.  
For HD\,7924, the measurement uncertainties (0.7\,\mse) and jitter (2.13\,\mse) 
predict the level of velocity rms (2.3\,\mse) in the absence of a planetary companion.
Thus, when interpreting this FAP, 
we should discount it by the probability that the observed velocity variability is due to noise.  

On balance, we reject the null hypothesis and with confidence attribute the observed 
velocity variation to a planet orbiting HD\,7924 with a period of 5.398\,d.

\section{Possible Second Companion}
\label{sec:second_planet}

\begin{figure}
\vspace*{0.25in}
\hspace*{0.1in}
\plotone{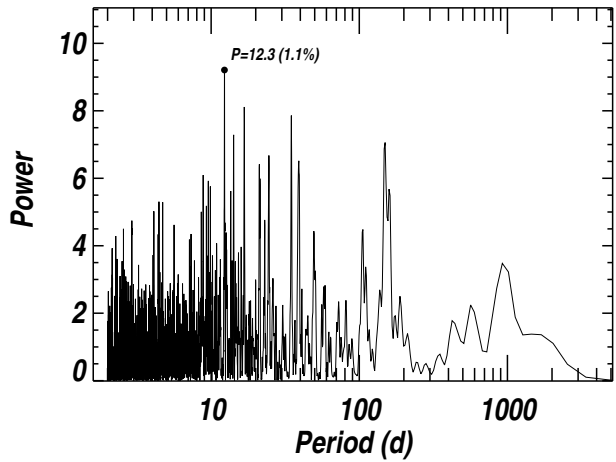}
\caption{Lomb-Scargle periodogram of velocity residuals to the single-planet fit show in Fig.\ \ref{fig:phased_1p}.  The peaks at 12.3, 16.7, 34.8, and 147.5 days suggest a 
possible second planet at one of these periods.  
Additional measurements are needed to tease out such a weak signal, if present.}
\label{fig:pergram_1p_resid}
\end{figure}

Motivated by the success of our single-Keplerian fit 
and the higher than expected velocity rms to that fit (2.78\,\mse), 
we performed a search for a double-Keplerian fit.
The Lomb-Scargle periodogram of the residuals to the single-planet fit 
(Fig.\ \ref{fig:pergram_1p_resid}) shows several short-period narrow peaks with 
modest power (at 12.3, 16.7, 34.8\,d) and an intermediate-period peak 
with broad power (at 147.5\,d). 
Each of these peaks represent a possible second planet in the system.  

We seeded the Keplerian search with several trial two-planet solutions.  
The first planet was always seeded with the orbital parameters 
for the best-fit single-planet model in \S\,\ref{sec:orbital}.
The second planet was seeded with the largest peaks in the periodogram 
of single-planet residuals (Fig.\ \ref{fig:pergram_1p_resid}) and a wide variety 
of other trial periods.  
All of the parameters were allowed to vary during the fitting process \citep{Wright08}.  
None of these two-planet fits changed the orbital parameters for HD\,7924b substantially.
The top four two-planet solutions had best-fit second-planet 
periods of 12.3, 34.8, 16.7, and 147.5\,d, 
and best-fit eccentricities of 0.72, 0.76, 0.44, 0.83, respectively.  
The high eccentricities raise our suspicion that the signals may be due to noise and 
increase the need for a robust tests of the null hypothesis.

We calculated FAPs for each of these two-planet models,
building on the methods described in \S\,\ref{sec:orbital}.
Synthetic data sets were constructed by drawing with replacement 
from the residuals to the best-fit single-planet solution and 
adding the coherent, best-fit solution back to the scrambled residuals.  
We performed a thorough search for the best-fit two-Keplerian model (as above)
on each synthetic data set.
False alarms were triggered when a synthetic data set had a lower value of \chinu\ 
than the original, unscrambled data.  
All of the two-planet models had an FAP\,$>$\,20\%, 
even when we restricted the scrambled trials to low-eccentricity solutions.  
Thus, we do not consider any of these signals to be viable planet candidates at the present time.
Nevertheless, we will continue to hunt for additional planets orbiting HD\,7924 by making 
additional observations and by refining our Doppler analysis algorithms to 
reanalyze our extant Doppler spectra more precisely.

\section{Photometric Observations}
\label{sec:photometry}

We acquired photometric observations of HD\,7924 with the T12 0.8\,m automated
photometric telescope (APT) at Fairborn observatory in southern Arizona.  
Our brightness measurements were made between 2006 December and 2008 October
and cover the last part of the 2006--07, the complete 2007--08, and the 
first part of the 2008--09 observing seasons.  The T12 APT and its precision
photometer are very similar to the T8 APT described in \citet{Henry1999}. The
precision photometer uses two temperature-stabilized EMI 9124QB 
photomultiplier tubes to measure photon count rates simultaneously through 
Str\"omgren $b$ and $y$ filters. 

The telescope was programmed to gather observations of HD\,7924 with respect 
to three nearby comparison stars in the following sequence:  DARK, A, B, C, 
D, A, SKY$_{\rm A}$, B, SKY$_{\rm B}$, C, SKY$_{\rm C}$, D, SKY$_{\rm D}$, 
A, B, C, D.  The comparison stars A, B, and C are HD\,4295 ($V=6.39$, 
$B-V=0.42$, F3~V), HD\,10971 ($V=6.94$, $B-V=0.49$, F5), and HD\,10259 
($V=6.59, B-V=1.04$, G5), respectively, while star D is HD\,7924 ($V=7.17, 
B-V=0.83$, K0).  

Each complete sequence, referred to as a group observation, was 
reduced to form 3 independent measures of each of the 6 differential 
magnitudes D$-$A, D$-$B, D$-$C, C$-$A, C$-$B, and B$-$A.  The differential 
magnitudes were corrected for differential extinction with nightly extinction 
coefficents and transformed to the standard Str\"omgren system with
yearly mean transformation coefficients.  The three independent measures 
of each differential magnitude were combined, giving one mean data point 
per complete sequence for each of the 6 differential magnitudes.  To filter
any observations taken under non-photometric conditions, an entire group
observation was discarded if the standard deviation of any of the six mean 
differential magnitudes exceeded 0.01 mag.  Finally, we combined the 
Str\"omgren $b$ and $y$ differential magnitudes into a single $(b+y)/2$ 
passband to improve the precision.

Our complete data set consists of 192, 435, and 145 good group observations 
from the 2006--07, 2007--08, and 2008--09 observing seasons, respectively,
for a total 772. To minimize the effect of any low-level intrinsic variation 
in the three comparison stars, we averaged the three D$-$A, D$-$B, and D$-$C 
differential magnitudes of HD\,7924 within each group into a single value, 
representing the difference in brightness between HD\,7924 and the mean of 
the three comparison stars:  $D-(A+B+C)/3$.  The standard deviations of 
these ensemble differential magnitudes for each of the three observing 
seasons are 1.77, 1.85, and 1.47\,mmag, respectively.  The standard
deviation of the complete data set is 1.76\,mmag.  These values are 
comparable to the typical precision of a single observation with our 
0.8\,m APTs \citep{Henry1999}, indicating there is little or no photometric 
variability in HD\,7924 or in the three comparison stars.  Solar-type stars 
often exhibit brightness variations caused by cool, dark photospheric spots 
as they are carried into and out of view by stellar rotation 
\citep[e.g.,][]{Gaidos2000}.  However, periodogram analysis of the individual 
three seasons of HD\,7924 and of the three observing seasons taken together, 
confirms the lack of any photometric periodicity between 1 and 100 days.  
This is consistent with the star's modest level of chromospheric 
activity and its low $v$\,sin\,$i$.

\begin{figure}
\epsscale{1.1}
\vspace*{0.05in}
\plotone{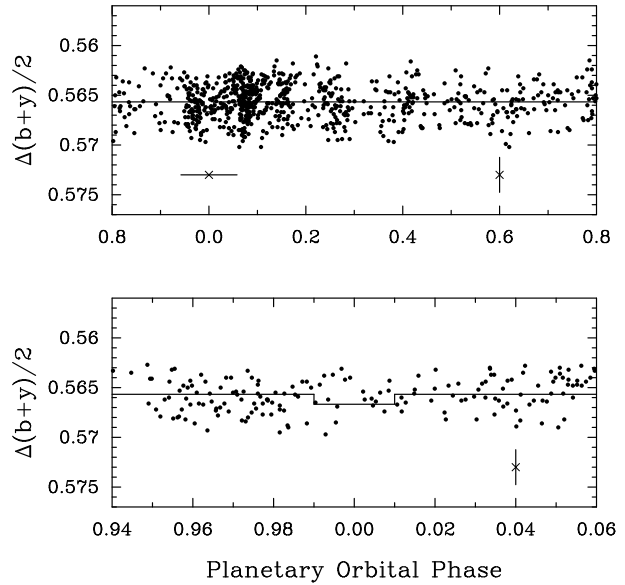}
\caption{$Top~Panel$: The 772 ensemble $D-(A+B+C)/3$ photometric 
observations of HD\,7924 in the combined Str\"omgren $(b+y)/2$ passband, 
acquired with the T12 0.8\,m APT over three observing seasons and plotted
modulo the 5.3978\,d orbital period of the super-Earth companion.  Phase 0.0 
corresponds to the predicted time of mid transit.  A least-squares sinusoidal fit 
at the orbital period yields a semi-amplitude of only $0.19\,\pm\,0.09$\,mmag.  
$Bottom~Panel$:  The photometric observations of HD\,7924 near the 
predicted time of transit replotted with an expanded scale on the abscissa.
The solid curve shows the predicted time of transit with a drop in stellar
brightness of 1\,mmag.  The error bar in the lower right of both 
panels represents the mean precision of a single observation (1.7\,mmag).  
The error bar immediately below the transit in the top panel represents the 
uncertainty in the predicted time of mid-transit ($\pm0.06$ phase units). 
The width of the plot in the bottom panel corresponds to this uncertainty in
the transit timing.
\label{fig:photometry}}
\end{figure}

The 772 ensemble $(b+y)/2$ differential magnitudes of HD\,7924 are plotted 
in the top panel of Fig.~\ref{fig:photometry}.  Phases are computed from 
the Keplerian orbital period 
$P$\,=\,5.3978\,$\pm$\,0.0015\,d (Table~\ref{tab:orbital_params})
and the epoch JD~2,454,727.99\,$\pm$\,0.16, 
a recent time of mid-transit derived from the orbital elements.  
The standard deviation of 
a single observation from the mean of the entire data set is 1.76\,mmag.
A least-squares sinusoidal fit on the orbital period yields a semi-amplitude of
only 0.19\,$\pm$\,0.09\,mmag.  This very low limit to photometric 
variability on the radial velocity period is strong evidence that the 
low-amplitude radial velocity variations observed in the star are, in fact, 
due to planetary reflex motion and not to activity-induced intrinsic 
variations in the star itself \citep[e.g.,][]{Paulson2004}.  

The photometric observations of HD\,7924 near the predicted time of transit 
are replotted with an expanded horizontal scale in the bottom panel of 
Fig.~\ref{fig:photometry}.  The solid curve shows the predicted time (0.00 phase units) and 
duration ($\pm$0.01 phase units) of transits with an arbitrary drop of 
1\,mmag from the mean brightness level of all the observations.  The error 
bar in the lower right of both upper and lower panels represents the mean 
precision of a single observation (1.7\,mmag).  The horizontal error bar 
immediately below the transit in the top panel represents the 2-$\sigma$ uncertainty in 
the predicted time of mid-transit ($\pm$0.32 days or $\pm$0.06 phase units). 
The \textit{a priori} probability of HD\,7924b transiting its host star is 0.06.
As discussed in \S\,\ref{sec:discussion}, the predicted depth for a transit of HD\,7924b ranges 
from 6\,mmag down to 0.3\,mmag, depending on the composition of the planet. 
Transit depths of 2--3 mmag or more are easily excluded by the data.  
The 20 observations within the nominal transit window between phases 0.99 and 
1.01 have a mean brightness just 0.06\,mmag fainter than the mean light 
curve level.  To estimate the limiting depth of detectable transits from our 
current data set, we segregated the observations betwen phase 0.93 and 0.07 
into seven bins each 0.002 phase units in width.  We computed the mean depth 
of each bin and compared them to the mean brightness level of all the 
observations.  The differences ranged from 0.46\,mmag fainter to 1.12\,mmag 
brighter than the overall mean brightness.  The standard deviation of the 
seven difference values is 0.4\,mmag.  Thus, we conclude 
that transits deeper than 2--3\,mmag do not occur, but additional 
observations around predicted transit times will be required to improve the 
lower limit or confirm transit depths around 0.5\,mmag or less.

\section{Discussion}
\label{sec:discussion}

We present the detection of HD\,7924b, a super-Earth planet with $M_P\sin i$\,=\,9.26\,$M_\earth$ 
in a $P$\,=\,5.398\,d orbit around a K0 dwarf.  This paper is the first in a series of 
super-Earth (\msinie\,$<$\,10\,$M_\earth$) and Neptune-mass 
planet announcements from the NASA-UC Eta-Earth Program.  
The 230 stars in this survey are nearby ($d$\,$<$\,25\,pc), 
bright enough for $\sim$1\,\ms RV measurements with Keck-HIRES 
($V$\,$<$\,11; some faint stars suffer from slightly poorer Doppler precision), 
have low to moderate chromospheric activity (\lrphk\,$<$\,$-4.7$), 
and are within two magnitudes of the main sequence.  
Several candidate super-Earth and Neptune-mass planets are emerging from the survey.
Some of these require additional follow-up observations to verify 
the orbit and accurately estimate Keplerian parameters.
Once the survey and follow-up observations are complete, 
the statistics of low-mass planets will offer strong constraints on planet-formation theory.
Just as successful theories of planet formation must reproduce the observed planet distribution 
in mass-period space and other parameterizations
for Jupiter- and Saturn-mass planets \citep{Cumming08}, they must do the same for low-mass planets.
In this regard, HD\,7924b is an interesting first detection for the Eta-Earth survey.
Its minimum mass of 9\,$M_\earth$ places it clearly in the ``planet desert'', 
while its small semi-major axis (0.057\,AU) pushes it into the 
pileup of short-period ``hot super-Earths'' and ``hot Neptunes'' seen 
in some simulations \citep{Ida_Lin08_iv}.
The existence of this population depends crucially on an unknown mechanism stopping 
Type I migration and parking the planets in short-period orbits.
The statistical power of the full Eta-Earth survey is needed to check whether 
HD\,7924b (as well as other super-Earths listed in Table~\ref{tab:super_earths}) 
is a member of this hypothesized class of planets 
or an outlier in a mostly barren patch of planet parameter space.

HD\,7924b is one of only eight super-Earth planets detected by the radial velocity technique 
and published (or submitted) in the refereed literature (Table~\ref{tab:super_earths}).  
The other planets are: 
HD\,40307b/c/d, the triple super-Earth system \citep{Mayor09}; 
GJ\,581c/d, the two planets in or near the habitable zone \citep{Udry2007}; 
GJ\,876d, the super-Earth in the same system as two resonant, 
intermediate-period Jupiters \citep{Rivera2005}; 
and GJ\,176b, an 8\,$M_\earth$ super-Earth \citep{Forveille09}.  

Table~\ref{tab:super_earths} reveals some emerging properties of super-Earths, 
albeit with the limited statistical power of only eight detections 
and through the lens of possible observational selection effects.
Only three of these planets have periods longer than 10\,d and only one has a period longer 
than 30\,d.  
All of these planets are in low eccentricity orbits, consistent with circular. 
As with higher-mass planets in short-period orbits, this is probably due to tidal circularization.
It is curious that  HD\,7924b and GJ\,176b are the only Doppler planets in this mass range 
\textit{not} in known multi-planet systems.  
This is quite different from the 14\% of higher-mass planets 
in known multi-planet systems \citep{Wright08b}.

\begin{deluxetable}{lccclcc}
\tabletypesize{\footnotesize}
\tablecaption{Planetary and Stellar Parameters for RV-Detected Super-Earths 
                        ($M_{\mathrm{pl}}$\,sin\,$i$ $\le$ 10\,\mearthe)
\label{tab:super_earths}}
\tablewidth{0pt}
\tablehead{
\colhead{Planet}   & 
\colhead{$M_{\mathrm{pl}}$\,sin\,$i$} &
\colhead{$P$}  &
\colhead{$a$}  &
\colhead{Spec.}  & 
\colhead{$M_{\star}$}& 
\colhead{\feh} \\
& 
(\mearthe) & 
(d) &
(AU)
& 
($M_{\odot}$) &
&
}
\startdata
HD\,7924b     &   9.3  &  \phn5.4  &  0.057     &   K0\,V      &  0.83    & $-$0.15      \\
HD\,40307b   &  4.2   &  \phn4.3  &  0.047     &  K2.5\,V   &  0.77    & $-$0.31      \\
HD\,40307c   &  6.9   &  \phn9.6  &  0.081     &  K2.5\,V   &  0.77    & $-$0.31      \\
HD\,40307d  &  9.2   &  20.5       &  0.134      &   K2.5\,V  &  0.77    & $-$0.31      \\
GJ\,176b       &  8.4   &  \phn8.8  &  0.066      &   M2\,V     &  0.53    & $-$0.1\phn \\
GJ\,581c       &  5.0   &  12.9       &  0.073       &   M3\,V     &  0.31    & $-$0.33      \\
GJ\,581d       &  7.7   &  83.6       &  0.25\phn &   M3\,V     &  0.31    & $-$0.33      \\
GJ\,876d       &  5.7   &  \phn1.9  &  0.021      &   M4\,V     &  0.32    & $-$0.12      \\
\enddata
\tablecomments{Planetary parameters (minimum mass, orbital period, and semi-major axis) 
of known super-Earth planets ($M_{\mathrm{pl}}$\,sin\,$i$ $\le$ 10\,\mearthe)
and stellar parameters (spectral type, mass, and iron abundance ratio) of their host stars.
Only planets detected by the RV method are shown. 
HD\,40307b/c/d were discovered by \citet{Mayor09}, 
GJ\,176b by \citet{Forveille09}, 
GJ\,581c/d by \citet{Udry2007}, 
and GJ\,876d by \citet{Rivera2005}.}
\end{deluxetable}

The host stars of known super-Earths also show emerging trends.
While Jovian-mass planets are found preferentially around 
higher-mass stars \citep{Johnson07}, 
the preliminary statistics of super-Earths are consistent with the opposite trend.  
The known super-Earth host stars are all K and M dwarfs in the mass range 0.31--0.83\,$M_{\odot}$.  
Current Doppler surveys are capable of detecting super-Earths 
orbiting G dwarfs (with reduced sensitivity), 
but none have been reported in the literature as of 2008 October.  
Note that super-Earths around most F stars and 
intermediate-mass sub-giants with $M_{\star}$\,$\gtrsim$\,1.3\,$M_{\odot}$ (``retired'' A stars) 
are undetectable by the Doppler method due to  
higher intrinsic RV variability observed in these targets.

The metallicities of super-Earth host stars also show the opposite trend seen in higher-mass stars.  
Note that \textit{all} of the detected super-Earths orbit metal-poor stars, 
in sharp contrast to the positive correlation observed between host star metallicity and 
Jupiter-mass planet occurance \citep{Fischer2005}.  

Additional planets and non-detections from unbiased stellar samples are needed to 
determine if the above trends reflect physical processes or selection effects.  
While planet detectability depends on many factors including 
the planet's Doppler semi-amplitude $K$, 
the amplitude of systematic and astrophysical errors, measurements precision, 
the number and timing of observations, the number of planets in the system, etc.,
we can say roughly that the lowest-mass detectable planets have $K$ 
comparable to the velocity rms of long-term stable stars (see Fig.~\ref{fig:standard_stars}).
Since the Doppler semi-amplitude of super-Earths scales as
\begin{equation}
K = \frac{3.7 \mathrm{\,m\,s}^{-1}}{(1-e^2)^{1/2}}
      \left( \frac{P}{5\mathrm{\,d}} \right) ^{-1/3} 
      \left( \frac{M_{\star}}{M_{\odot}} \right) ^{-2/3} 
      \frac{M\mathrm{_{pl}\,sin\,}i}{10\mathrm{\,}M_\earth}, 
\end{equation}
RV surveys are biased against low-mass, long-period planets orbiting 
high-mass stars.  
For comparision, the known super-Earths listed in Table~\ref{tab:super_earths} 
have $K$\,=\,2.0--6.5\,\mse.

HD\,7924 is nearby and bright, making plausible the detection of its short-period planet 
by other techniques.
While our photometric observations preclude a transit down to the level of approximately 0.5\,mmag 
during the times described in \S\,\ref{sec:photometry}, 
and the \textit{a priori} transit probability for HD\,7924b is 0.06, 
follow-up campaigns are warranted given the extraordinary value of a nearby transiting super-Earth.
Further, while the radius of HD\,7924b is unknown, we can place bounds of 
$R_{\mathrm{pl}}$\,=\,1.4--6\,\rearth for planet models spanning 
pure iron to pure hydrogen \citep{Seager2007}.  
These extremes produce transits of depth 0.3 and 6\,mmag, respectively.  
A more plausible silicate-water planet with $R_{\mathrm{pl}}$\,$\approx$\,2\,\rearth  
\citep{Seager2007} produces a $\sim$0.6\,mmag transit, near our detection limit.

With an astrometric signature of 0.5\,$\mu$as, 
detecting HD\,7924b astrometrically will only be possible with 
NASA's SIM (Space Interferometry Mission) or its proposed variants 
(e.g.\ Sim-Lite and Planet Hunter) that achieve a single measurement precision 
of $\lesssim$\,1\,$\mu$as.  
Additional planet(s) of comparable mass in larger orbits, should they exist,  
would presumably be easier to detect.

Direct imaging of HD\,7924b is infeasible for the foreseeable future given the 
maximum projected separation of $\sim$15\,mas for this short-period planet.

As Doppler velocity surveys probe the observationally challenging realm 
of super-Earths and Neptune-mass planets, robust detections require 
continued improvement in velocity precision, an increased number 
of observations per target, and extra care interpreting the velocities
with careful consideration of the null hypothesis.  Because the Doppler 
velocity amplitudes for these lowest mass planets are similar in 
magnitude to the single measurement precision, we are in an era 
where it is important to demonstrate the fundamental velocity precision 
and long term instrumental stability, and to provide velocity data so that 
others may check the confidence level of the claim and explicitly 
consider the null hypothesis by carrying out FAP estimates and other tests. 
These safeguards will ensure continued confidence and credibility 
of the profound and extraordinary claims of the detection of low-mass planets.

\acknowledgments{We gratefully acknowledge the efforts and dedication of the 
Keck Observatory staff.  
We are also grateful to the time assignment committees of NASA, NOAO, and 
the University of California for their generous allocations of observing time.  
We acknowledge R.\ Paul Butler and S.\ S.\ Vogt for many years
of contributing to the data presented here.
A.\,W.\,H.\ gratefully acknowledges support from a Townes Post-doctoral Fellowship 
at the U.\,C.\ Berkeley Space Sciences Laboratory.
J.\,A.\,J.\ is an NSF Astronomy and Astrophysics Postdoctoral Fellow and
acknowledges support form NSF grant AST-0702821.
G.\,W.\,M.\ acknowledges NASA grant NNX06AH52G.  
G.\,W.\,H.\ acknowledges support from NASA, NSF, Tennessee State University, and
the State of Tennessee through its Centers of Excellence program.
Finally, the authors wish to extend special thanks to those of Hawaiian ancestry 
on whose sacred mountain of Mauna Kea we are privileged to be guests.  
Without their generous hospitality, the Keck observations presented herein
would not have been possible.}

\bibliographystyle{apj}
\bibliography{ms}


\LongTables  
\begin{deluxetable}{ccc}
\tabletypesize{\footnotesize}
\tablecaption{Radial Velocities for HD\,7924
\label{tab:keck_vels}}
\tablewidth{0pt}
\tablehead{
\colhead{}         & \colhead{Radial Velocity}     & \colhead{Uncertainty}  \\
\colhead{JD -- 2440000}   & \colhead{(\mse)}  & \colhead{(\mse)}  
}
\startdata
 12187.970 &    5.62 &    1.3  \\
 12218.998 &    2.86 &    1.3  \\
 12220.015 &    2.95 &    1.7  \\
 12307.770 &    9.57 &    1.3  \\
 12535.955 &    7.59 &    1.2  \\
 12575.915 &   -2.55 &    1.6  \\
 12850.115 &    1.19 &    1.3  \\
 13239.082 &   -2.93 &    0.9  \\
 13338.794 &    3.20 &    0.9  \\
 13400.804 &   -1.90 &    0.7  \\
 13692.825 &   -0.08 &    0.5  \\
 13693.006 &   -0.09 &    0.5  \\
 13693.777 &   -0.40 &    0.5  \\
 13693.952 &   -1.26 &    0.5  \\
 13694.820 &    0.87 &    0.5  \\
 13694.956 &    2.40 &    0.5  \\
 13695.010 &    1.45 &    0.6  \\
 13695.898 &    5.29 &    0.3  \\
 13696.929 &    2.99 &    0.5  \\
 13724.850 &   -4.04 &    0.7  \\
 13746.799 &    3.51 &    0.7  \\
 13748.790 &    3.77 &    0.6  \\
 13749.785 &    5.47 &    0.6  \\
 13750.795 &    1.05 &    0.6  \\
 13751.831 &   -1.53 &    0.7  \\
 13752.836 &   -6.05 &    0.7  \\
 13753.801 &   -3.46 &    0.6  \\
 13775.767 &    3.38 &    0.7  \\
 13776.805 &    7.31 &    0.6  \\
 13777.805 &    4.78 &    0.7  \\
 13778.799 &   -0.62 &    0.7  \\
 13779.823 &   -3.30 &    0.7  \\
 13933.124 &    6.05 &    0.6  \\
 13959.130 &    0.26 &    0.6  \\
 13961.070 &    4.64 &    0.5  \\
 13962.067 &    2.93 &    0.5  \\
 13963.104 &   -3.91 &    0.6  \\
 13964.134 &   -0.83 &    0.9  \\
 13982.026 &    7.67 &    0.5  \\
 13983.083 &    5.20 &    0.6  \\
 13983.987 &   -2.50 &    0.6  \\
 13985.005 &    0.89 &    0.6  \\
 14083.895 &    8.07 &    0.7  \\
 14084.837 &   10.80 &    0.7  \\
 14129.917 &   -0.30 &    0.7  \\
 14130.714 &   -1.69 &    0.7  \\
 14131.780 &    0.76 &    0.8  \\
 14138.764 &    1.39 &    0.7  \\
 14295.133 &    2.63 &    0.7  \\
 14305.135 &    3.12 &    0.8  \\
 14306.136 &    2.58 &    0.7  \\
 14307.122 &   -0.45 &    0.6  \\
 14308.117 &   -6.29 &    0.9  \\
 14309.108 &   -6.06 &    0.6  \\
 14310.108 &   -0.86 &    0.6  \\
 14311.093 &    1.29 &    0.6  \\
 14312.089 &    3.27 &    0.6  \\
 14313.084 &   -3.57 &    0.7  \\
 14314.089 &   -2.89 &    0.7  \\
 14315.106 &    1.35 &    0.6  \\
 14319.125 &   -6.39 &    0.5  \\
 14336.084 &   -7.84 &    0.7  \\
 14337.143 &   -1.27 &    1.0  \\
 14339.112 &    2.83 &    0.9  \\
 14344.047 &    2.72 &    0.7  \\
 14345.147 &   -5.63 &    0.8  \\
 14396.864 &   -0.96 &    1.0  \\
 14397.921 &    3.20 &    0.9  \\
 14398.932 &    1.66 &    1.0  \\
 14399.877 &    4.37 &    1.0  \\
 14427.831 &    2.43 &    0.6  \\
 14429.895 &    5.23 &    0.7  \\
 14430.934 &    3.91 &    0.8  \\
 14453.738 &   -1.64 &    0.9  \\
 14454.786 &   -4.08 &    0.7  \\
 14455.788 &   -3.85 &    0.6  \\
 14456.771 &    0.00 &    0.8  \\
 14458.922 &  -10.72 &    1.3  \\
 14460.768 &   -2.41 &    0.9  \\
 14461.695 &   -0.77 &    1.9  \\
 14492.794 &   -3.44 &    0.5  \\
 14675.111 &   -3.67 &    0.6  \\
 14690.113 &   -5.83 &    0.7  \\
 14718.021 &   -4.44 &    0.6  \\
 14719.063 &   -2.22 &    0.5  \\
 14720.076 &   -0.94 &    0.6  \\
 14721.069 &    3.44 &    0.5  \\
 14722.117 &   -1.11 &    0.6  \\
 14723.038 &   -2.72 &    0.6  \\
 14724.006 &   -3.75 &    0.6  \\
 14725.049 &   -0.92 &    0.6  \\
 14727.066 &    2.59 &    0.5  \\
 14727.989 &   -0.71 &    0.5  \\
\enddata
\end{deluxetable}

\enddocument